\documentclass[letter]{IEEEtran}
\usepackage{graphicx}
\usepackage{amsmath}
\usepackage{subfigure}
\usepackage{cite}

\ifCLASSINFOpdf

\else

\fi

\begin{document}

\title{Histogram Transform-based Speaker Identification}

\author{Zhanyu~Ma, Hong~Yu
\thanks{Z. Ma and H. Yu are with Pattern Recognition and Intelligent System Lab., Beijing University of Posts and Telecommunications, Beijing, China.}
}

\maketitle

\begin{abstract}
A novel text-independent speaker identification (SI) method is proposed.
This method uses the Mel-frequency Cepstral coefficients (MFCCs) and
the dynamic information among adjacent frames as feature sets to capture speaker's characteristics. In order to utilize dynamic information,
we design super-MFCCs features by cascading three neighboring MFCCs frames together.
The probability density function (PDF) of these super-MFCCs features is estimated
by the recently proposed histogram transform~(HT) method, which generates more training data
by random transforms to realize the histogram PDF estimation and recedes the commonly occurred
discontinuity problem in multivariate histograms computing. Compared to the conventional PDF
estimation methods, such as Gaussian mixture models, the HT model shows promising improvement in the SI performance.
\end{abstract}

\IEEEpeerreviewmaketitle
\section{Introduction}
\label{sec:introduction}

\IEEEPARstart{S}{peaker} identification is a biometric task that has been intensively studied
in the past decades~\cite{S2012,ZMA2011,ZMA2013,Y2013,Ma2012}.~Given an input speech,
the task is to determine the unknown speaker's identity by selecting one speaker from the whole set
of speakers registered in the system~\cite{Y2013,Ma2010}.

The first step is feature extraction.
In this part the original speech signals are transformed into feature vectors which can represent speaker-specific properties.~To this end, a lot of features have been considered,~\emph{e.g.},~the Mel-frequency Cepstral coefficients (MFCCs)~\cite{Sahidullah2011}, and the line spectral frequencies (LSFs)~\cite{ZMA2011}.
Among them, MFCCs are widely used in speech processing tasks,~\emph{e.g.},~language identification~\cite{U2013}, speech emotion classification~\cite{D2013}, and speaker identification~\cite{T2005}. In general, these static features
are supplemented by their corresponding velocity and acceleration coefficients to get dynamic information. Recently, some researchers tend to use the static features to directly build the system. In~\cite{ZMA2011,ZMA2013,Ma2010a}, LSFs are directly used in super-Dirichlet mixture models and in~\cite{Pan2012,Ma2009,Ma2008}, static MFCCs are used in the deep learning model. In this paper, we also adopt the static MFCCs feature and, moreover, group several neighboring frames together to create a super MFCCs feature to express the speaker's characteristics.

The second step is model training. As the extracted features can describe the unique characteristic of
an individual speaker, this allows us to classify each speaker by their voices using probabilistic
models~\cite{Manas2013}. Separate models should be trained for each speaker,
in order to describe the statistical properties of the extracted features.

The third step is identification. In this part the feature vectors extracted from the unknown person's
speech are compared against the models trained in the second step to make the final decision by using the maximum likelihood method.

The effectiveness of a speaker identification system is mainly decided by the design of the statistical model in
the second part. The mixture model based methods are widely employed, e.g., Dirichlet mixture model (DMM)~\cite{ZMA2011,zMa2014,zMa2014pr}, beta mixture model (BMM), von-Mises Fisher mixture model~\cite{J2013,J2014} and Gaussian mixture model (GMM)~\cite{D1995,S2004,Ma2014c,Ma2015a}. All these models belong to parametric models, where the aim of training is to optimize the parameters of the models.

In addition to the mixture model based approaches, nonparametric approaches which can offer close adaptation to particular
features of the training data are also widely used~\cite{J1994,W2004}. One of the most popular non-parametric approaches is the histogram probability estimation. Partitioning the training feature space into discrete intervals (bins), we can get the probability estimation by counting the number of training data that fall into each bin. If we have sufficient training data and set an appropriate bin width, good performance can be obtained~\cite{WN2002}. However, probability estimated by the histogram method, especially the multivariate histograms-based method, has large discontinuities~\cite{E2014}. This is because the bin number will increase at a geometrical ratio with the growing of the feature's dimensionality. When the dimensionality is high, we can't get sufficient training data in order to cover all the bins in the space. Recently, a histogram transform (HT) model was proposed to overcome such problems~\cite{E2014}. The HT model can alleviate the discontinuity problem by averaging multiple multivariate histograms. This method has been successfully applied in several applications, such as image segmentation~\cite{E2014}. In this paper, we will use this method to build speaker identification models.

In the experimental part, we compare the performance of the HT model with the GMM model.
The identification decision was made by choosing the maximal log-likelihood of a test speech against all the trained speaker models. Experimental results show that the HT model is able to reach higher accuracy than the GMM model. This paper is organized as follows: The way to generate the super MFCCs features is described in Section~\ref{sec:Feature}. We describe the HT model in Section~\ref{sec:train}. The experimental results and analysis are presented in Section~\ref{sec:experiment}. Conclusions and some further work are given in Section~\ref{sec:conclusion}.

\section{Feature Extraction}
\label{sec:Feature}
In speech processing, the mel-frequency cepstrum~(MFC) is a representation of
the short-term power spectrum of a speech signal, based on a linear cosine transform of a log power spectrum on
a nonlinear mel-scale of frequency~\cite{Sahidullah2011}. MFCCs are coefficients that
collectively compose an MFC. They are derived from a type of cepstral representation
of the audio. For the speech segment~(frame) at time $t$, we can extract a $D$ dimensional MFCC vector as
\begin{equation}
\mathbf{x}{(t) = \left[x_1(t), \ldots, x_{\emph{D}}(t)\right]^{\rm{T}}}.
\end{equation}
In order to exploit the dynamic information useful for speaker recognition,
the traditional method is to construct a super feature vector containing the MFCCs, the velocity of the MFCCs ($\Delta {\bf{x}}(t)$),
and the acceleration of MFCCs ($\Delta \Delta {\bf{x}}(t)$)~\cite{J2008}. The super frame is then defined as
\begin{equation}
\Delta {\bf{MFC}}{{\bf{C}}_{\sup }}(t) = {[{\bf{x}}{(t)^{\rm{T}}},\Delta {\bf{x}}{(t)^{\rm{T}}},\Delta \Delta {\bf{x}}{(t)^{\rm{T}}}]^{\rm{T}}}.
\end{equation}

Inspired by the idea proposed and used in~\cite{ZMA2011,Pan2012,J2013,J2014}, we represent the dynamic information of the MFCCs in a new way. We build the super frame by utilizing two neighbors of the current frame, one from the past and the other from the following frames. Set the time interval between two adjacent frames as~$\tau$ and the super MFCCs frame ${\bf{x}}_{\sup}(t)$ is created by grouping the current frame and two neighbors as:
\begin{equation}
{{\bf{x}}_{\sup }}(t) = {[{\bf{x}}{(t - \tau )^{\rm{T}}},{\bf{x}}{(t)^{\rm{T}}},{\bf{x}}{(t + \tau )^{\rm{T}}}]^{\rm{T}}},
\end{equation}
where $\tau$ is an integer (\emph{e.g.}, $\tau=1$).
The conventionally used $\Delta {\rm{MFCC}}_{{\rm{sup}}}(t)$ feature contains processed information from the neighbor frames. The super MFCCs frame ${\bf{x}}_{\sup }(t)$ mentioned above actually includes the raw information contained in the neighbor frames. In principle, the super MFCCs frame ${\bf{x}}_{\sup }(t)$ should carry at least the same information as that represented by $\Delta {\rm{MFCC}}_{{\rm{sup}}}(t)$. This motivates us to use the ``raw'' data here.
\section{Training of the HT Models}
\label{sec:train}
Theoretically, the non-parametric probabilistic models, such as histogram
based models, are driven by training data directly and can simulate any complicated probability density function (PDF). In practice, the original histogram methods, especially the multivariate histograms-based methods, are rarely used due to the fact that the learned PDF has large discontinuities over the boundaries of the bins.

Fig.$2$~(a) shows the negative logarithm of PDF estimated for two randomly selected dimensions of $48$-dimensional ${\bf{x}}_{\sup }$ features using the original histogram method. The 16-dimensional MFCCs vectors $\mathbf{x}{(t)}$ are extracted from wide-band speech in the TIMIT dataset. The feature space is segmented into $40{\times}40$ bins and only $17.13{\%}$ zones have been filled and many zones yield zero (black color).
\begin{figure}[t]
\vspace{0mm}\centering
 \includegraphics[width=0.45\textwidth]{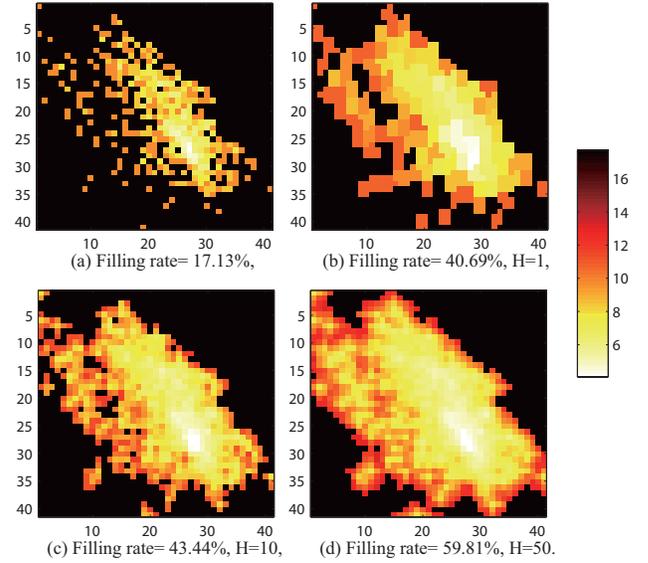}\\
 \vspace{0mm}
 \caption{
The original histogram and the transformed ones via HT. (a) The original one. (b) The histogram with one HT.
(c) The average histogram with 10 HTs. (d) The histogram with 50 HTs. Filling rate is the ratio of the number of non-zero bins to the total number of bins. H means the transform times. The values of the negative logarithm of PDF are plotted. The black color denotes zero density and white color presents the highest density.
  }
 \label{fig:HT}
  \vspace{0mm}
\end{figure}

In order to get a smooth PDF, parametric probabilistic models, such as mixture models, are usually employed. In these models, the combination of some simple smooth functions are recommended to estimate the actual PDF.
If the function form and number of mixture components are chosen appropriately, the mixture models can fit the real probability distribution well. However, when the actual PDF is overcomplex, the combination of several simple functions can not represent the true PDF properly.

Recently, an HT method was proposed in~\cite{E2014}.
The HT method applies a group of random affine transforms to the training data and computes the average histogram to estimate the PDF. As illustrated in Fig.$2$~(b), after one transform, some points fall in the zones where the original histogram has zero density and
$40.69{\%}$ bins have been filled. The PDFs estimated by the average histogram of 10 and 50 transforms are shown in Figs.$2$~(c) and (d), respectively. It is observed that the PDFs have been smoothed, the filling rates increase to $43.44{\%}$ and $59.81{\%}$, respectively, and the discontinuity has then been overcome.

The HT model is based on histogram methods, and it has advantage of strong adaptability. Meanwhile, the transformation can overcome the disadvantage of discontinuity. A parametric probability density function is adopted in this model as prior, so some merits of parametric models are also found in this method.

The affine function in the HT model is defined as
\begin{equation}
 {\bf{AF}}({\bf{x}};{\bf{A}},{\bf{b}}) = {\bf{Ax}} + {\bf{b}},
\end{equation}
where $\bf{x}$ is a training sample vector with size $D\times 1$, $\bf{A}$ is a $D\times D$ matrix, $\bf{b}$ is a $D\times 1$ vector. After $H$ times randomizing transforms, one training dataset
${\bf X}= [{{\bf{x}}_1}, \ldots,{{\bf{x}}_N}]$ of $N$ samples is mapped to $H$ training datasets. Then using the average histogram of these datasets to learn the PDF can partly solve the discontinuous problem~\cite{E2014}. Based on the above affine function incurred transforms, the probability of an input feature vector ${{\bf{x}}_{\text{in}}}$ in the HT method is defined as
\begin{equation}
\label{eq7}
{\mathop{\rm HT}\nolimits} ({{\bf{x}}_{\text{in}}};{ {\bf X}}) = {\pi_{0}}{\mathop{\rm P}\nolimits} ({{\bf{x}}_{\text{in}}}|{ {\bf X}})_0 + \frac{{1 - {\pi_{0}}}}{H}\sum\limits_{i = 1}^H {{\mathop{\rm P}\nolimits} ({{\bf{x}}_{\text{in}}}|{{\bf{A}}_i},{{\bf{b}}_i},{{\bf X}})}.\\
\end{equation}

The first item of~\eqref{eq7} is a priori probability of finding a test sample in a zone where the histograms yield zero density,\\${\pi_{0}}$ is defined as ${\pi_{0}}=(N+1)^{-1}$
and ${\mathop{\rm P}\nolimits} ({{\bf{x}}_{\text{in}}}|{{\bf X}})_0$ is defined as a multivariate Gaussian distribution,
\begin{equation}
{\mathop{\rm P}\nolimits} {(\left. {{{\bf{x}}_{{\rm{in}}}}} \right|{\bf{X}})_0} =\mathcal{N}({{\bf{x}}_{{\rm{in}}}};{\bf{\mu }},{\bf{C}}),
\end{equation}
\begin{equation}
{\bf{\mu }} = \frac{1}{N}\sum\limits_{j = 1}^N {{{\bf{x}}_j}}, {\bf{C}} = \frac{1}{{N - 1}}\sum\limits_{j = 1}^N {({{\bf{x}}_j} - {\bf{\mu }}){{({{\bf{x}}_j} - {\bf{\mu }})}^{\rm{T}}}}.
\end{equation}
The second item in~\eqref{eq7} describes the average histogram probability and ${\mathop{\rm P}\nolimits} ({{\bf{x}}_{\text{in}}}|{{\bf{A}}_i},{{\bf{b}}_i},{{\bf X}})$ is the histogram probability of the input data in the $i$-th transform. Following the method introduced in~\cite{E2014}, through adjusting the scale factor of the matrix $\bf{A}$, the bin width ${h^*}$ on the transformed space can be chosen as ${h^*} = 1$. Set
\begin{equation}
{{\bf{y}}_{i,\text{in}}} = {\mathop{\rm round}\nolimits} \left( {{\mathop{\rm \mathbf{AF}}\nolimits} ({{\bf{x}}_{\text{in}}};{{\bf{A}}_i},{{\bf{b}}_i})} \right),
\end{equation}
\begin{equation}
{{\bf{y}}_{ij}} = {\mathop{\rm round}\nolimits} \left( {{\mathop{\rm \mathbf{AF}}\nolimits} ({{\bf{x}}_j};{{\bf{A}}_i},{{\bf{b}}_i})} \right),
 \end{equation}
 where ${\mathop{\rm round}\nolimits}$ function means changing the components of the transformed vector to the nearest integer, then the histogram probability of input data ${{\bf{x}}_{\text{in}}}$ in the $i$-th transform is defined as
\begin{equation}
\label{eq8}
{\mathop{\rm P}\nolimits} ({{\bf{x}}_{\text{in}}}|{{\bf{A}}_i},{{\bf{b}}_i},{\bf{X}}) = \frac{1}{{N{v_i}}}\sum\limits_{j = 1}^N {{\mathop{\rm II}\nolimits} ({{\bf{y}}_{i, \text{in}}},{{\bf{y}}_{ij}})}.
\end{equation}
 In~\eqref{eq8}, ${{v_i}}$ is the D-dimensional volume of the histogram bins in the input space, as
\begin{equation}
{v_i} = |{{\bf{A}}_i}{|^{ - 1}}.
\end{equation}
${\mathop{\rm II}\nolimits}$ stands for the indicator function, defined as
\begin{equation}
{\mathop{\rm II}\nolimits} ({\bf{x}},{\bf{y}}) = \left\{ {\begin{array}{*{20}{c}}
1,&{{\bf{x}} = {\bf{y}}}\\
0,&{{\bf{x}} \ne {\bf{y}}}
\end{array}} \right..
\end{equation}
 The selection of the transform parameters ${\bf{A}}$ and ${\bf{b}}$ should take the following rules. Since the bin width on the transformed space is ${h^*=1}$, we draw ${\bf{b}}$ from the uniform distribution over the hypercube ${[0,1]^D}$.

 The matrix ${\bf{A}}$ can be expressed as the product of a unit rotation matrix ${\bf{U}}$ and a diagonal scaling matrix ${\bf{\Lambda }}$.
The random unit rotation matrix ${\bf{U}}$ can be generated by making QR decomposition on a standard normal random matrix~\cite{F2007}.

${\lambda _k}$, the diagonal elements of ${\bf{\Lambda }}$, can be generated using Jeffrey¡¯s prior for the scale parameters~\cite{H1946}.
$\log (\lambda {}_k)$ should be drawn from the uniform distribution over certain interval of real numbers $[\log ({\lambda _{\min }}),\log (\lambda {}_{max})]$, where
\begin{equation}
\log ({\lambda _{\min }}) = {\theta _{\min }} + \log (\hat \lambda ),
\end{equation}
\begin{equation}
\log ({\lambda _{\max }}) = {\theta _{\max }} + \log (\hat \lambda ).
\end{equation}
In order to make the bin width on the transformed space equal to $1$, according to the multivariate histograms theory~\cite{D1992}, $\hat \lambda$ should be set as
\begin{equation}
\hat \lambda  = \frac{{{N^{\frac{1}{{2 + D}}}}}}{{3.5}}\sqrt {\frac{D}{{trace({\bf{C}})}}}.
\end{equation}
${\theta _{\min }}$ and ${\theta _{\max }}$ are tunable parameters.~In this paper we empirically choose ${\theta _{\min }}=0$ and ${\theta _{\max }}=2$.

 \section{Experimental Results and Discussions}
\label{sec:experiment}
To verify the proposed HT model-based SI method, we evaluated the speaker identification performance on the TIMIT database~\cite{TIMIT1990}. The TIMIT database contains $630$ male and
female speakers coming from $8$ different regions and each
speaker spoke $10$ sentences. During each round of evaluation, we randomly selected $100$ speakers from the database.

The speech was segmented into frames with a $25$ms duration and a $10$ms step size. The silence frames were removed. For each frame, a Hanning window was used to reduce the high frequency components. Since the speech clips are wide band data, $16$-dimensional MFCCs were extracted from each frame. In order to compare the differences between the traditional $\Delta {\rm{MFCC}}_{{\rm{sup}}}$ and the super frame ${\bf{x}}_{\sup }$ proposed in this paper, the MFCCs and the corresponding velocity and acceleration features were calculated according to the methods described in Section~\ref{sec:Feature}. Finally, we obtained two sets of super frames, each contains $48$-dimensional features.

\begin{figure}[t]
\vspace{0mm}
\centering
 \subfigure[Using $\Delta {\rm{MFCC}}_{{\rm{sup}}}$ and ${\bf{x}}_{\sup }$ in the HT model.]{\includegraphics[width=0.45\textwidth]{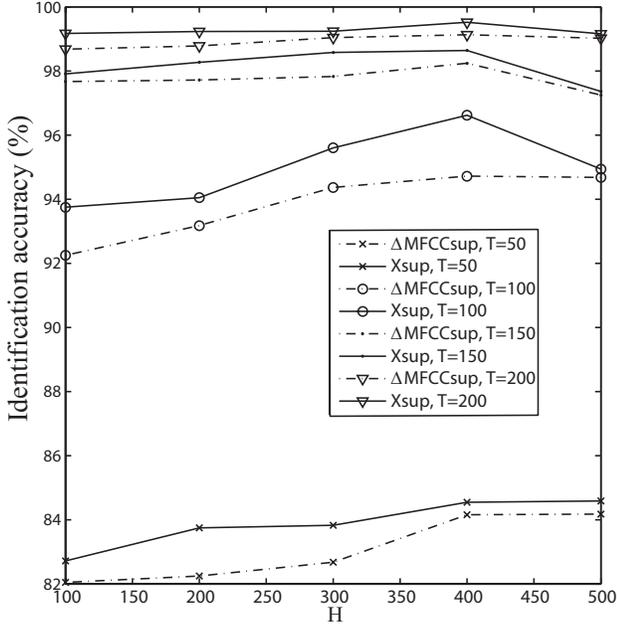}}
   \subfigure[Between the HT and GMM models.]{\includegraphics[width=0.45\textwidth]{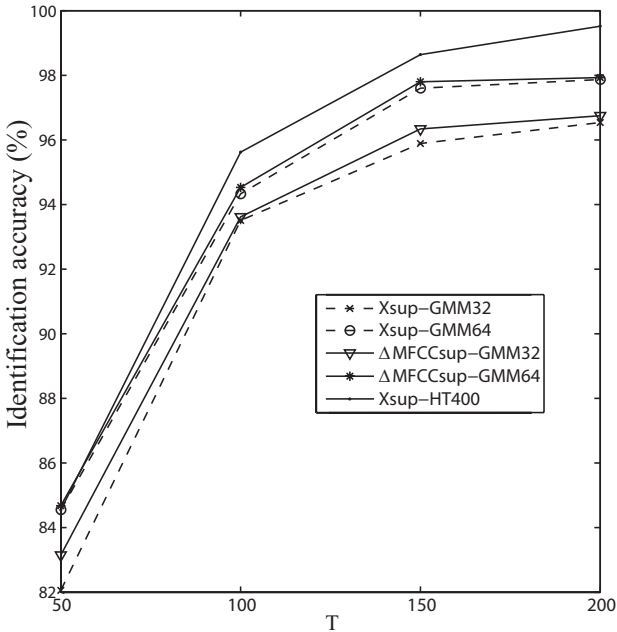}}
 \vspace{0mm}
\caption{Comparison of identification accuracies.}
\label{erroshow}
\vspace{0mm}
\end{figure}

In the training phase, seven sentences were randomly selected from each speaker as the training data and the remaining three sentences were used for testing. In each test sentence we randomly intercepted $10$ segments, each including $T$ consecutive frames, as test sets, so there were $3\times10\times100=3000$ test sets in total. We trained $100$ HT models using ${\rm{MFCC}}_{{\rm{sup}}}$ and ${\bf{x}}_{\sup }$ frames, respectively. Put a test segment into each trained model and the log-likelihood was calculated as:
\begin{equation}
\label{eq20}
{{\mathop{\rm L}\nolimits} _j}({\bf{\tilde X}}) = \sum\limits_{i = 1}^T {\log \left( {{\mathop{\rm HT}\nolimits} ({{\bf{x}}_i};{{\bf X}_j})} \right)},
\end{equation}
where ${\bf{\tilde X}}$ is the input segment set including $T$ feature frames, ${{{\bf{x}}_i}}$ denotes the $i$-th frame and ${{\bf X}_j}$ stands for the
training set of the $j$-th person. The trained model that yielded the largest log-likelihood value was considered to have the same statistical property as the test feature set, and therefore, we assigned the test segment with the identity of this trained model.
We set the number of transforms $H$ as \{$100, 200, 300, 400, 500$\} and the frame interval ${\tau=1}$. The frame number $T$ in each test set was chosen as \{$50, 100, 150, 200$\}, which means the durations of each test utterance is \{$0.5, 1, 1.5, 2$\} seconds, respectively. The identification score is calculated by the number of correctly identified test sets divided by the total number of test sets, we ran evaluation for $10$ rounds, and the average scores in different parameter and methods were shown in Fig. \ref{erroshow}.

The performance of using $\Delta {\rm{MFCC}}_{{\rm{sup}}}$ and ${\bf{x}}_{\sup } $ in HT model is shown in Fig.~\ref{erroshow}(a). It is observed that, the HT model with ${\bf{x}}_{\sup } $ frames reaches a higher identification accuracy. This indicates that the proposed ${\bf{x}}_{\sup } $ features are more suitable for the HT model than $\Delta {\rm{MFCC}}_{{\rm{sup}}}$. As introduced in Section~\ref{sec:train}, the data transform matrix $\bf{A}$ is generated according to a single parameter $\hat \lambda$ , so the feature ${\bf{x}}_{\sup }$ in which all components have similar attribute fits the HT model better.

The result also shows that the number of transforms $H$ affects the final score. Increasing $H$ improves the identification accuracy, but when $H$ is higher than $400$, the accuracy decreases instead. This indicates that too many transformations will make the estimated PDF over-smooth and with reduced speaker specific information. For example, when the speech duration is longer,~\emph{ e.g.}, more than $0.5s$, we have sufficient amount of feature frames to describe the speaker's characteristics, and less error caused by one frame can be compensated by the average of other frames. Hence, we want to increase the ``specificity'' of each frame, which means we want a ``cliffy'' PDF curve. Therefore, smaller $H$ is required in this case. However, when less amount of feature frames are presented, the requirement of smoothness get higher. Thus, larger $H$ should be employed in order to obtain a smooth PDF curve.

We also trained and tested the data sets in GMM models with different numbers of components,~\emph{i.e.}, $\{32,64\}$. The results are shown in Fig.~\ref{erroshow}(b). The $\Delta {\rm{MFCC}}_{{\rm{sup}}}$  features give better results in the GMM model. This means that the $\Delta {\rm{MFCC}}_{{\rm{sup}}}$  features are more suitable for the GMM model. This also verifies the well-known strategy utilized in SI tasks. Based on the above facts, ${\bf{x}}_{\sup}$ performs better in the HT model and $\Delta {\rm{MFCC}}_{{\rm{sup}}}$ better in the GMM model. When the number of test segments is relatively larger (e.g., more than 50 frames) the ${\bf{x}}_{\sup }$+HT methods can get lower error rates than the $\Delta {\rm{MFCC}}_{{\rm{sup}}}$+GMM method.

\begin{figure*}[!t]
\vspace{0mm}
\centering
 \includegraphics[width=0.85\textwidth]{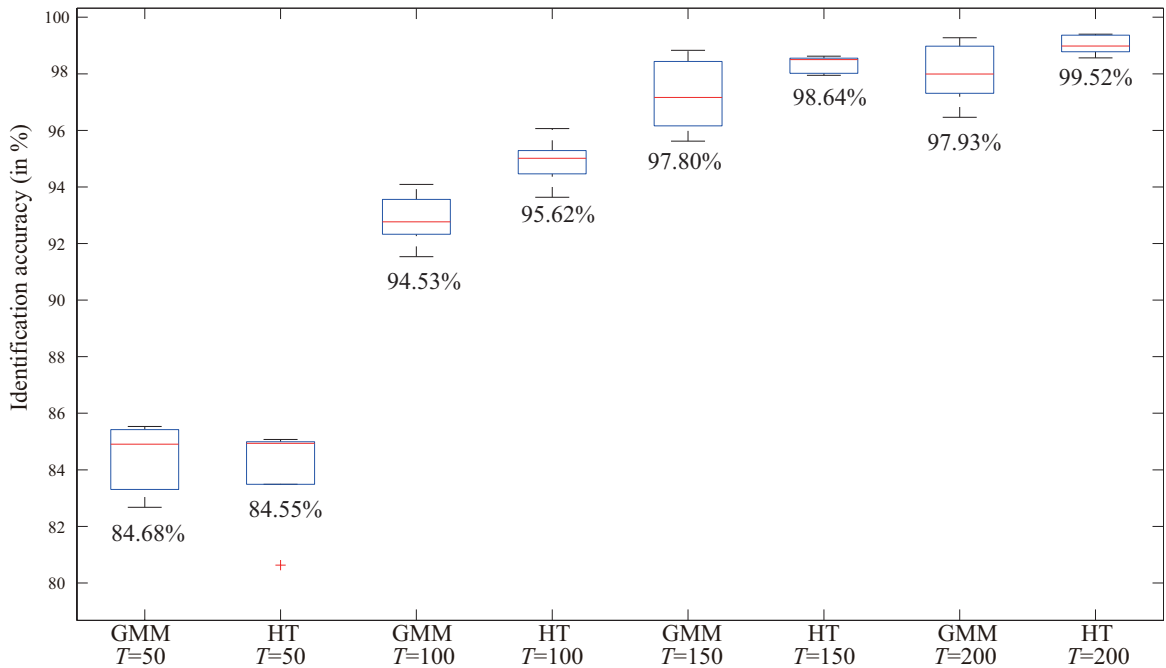}
 \vspace{0mm}
\caption{Comparison of the identification accuracy between GMM model with $64$ components  using $\Delta {\rm{MFCC}}_{{\rm{sup}}}$ features and HT model setting $H=400$ using $\Delta {\rm{MFCC}}_{{\rm{sup}}}$ features in different $T$. The central red mark is the median,the edges of the box are the $25^{th}$ and $75^{th}$ percentiles. The outliers are marked with red crosses and the mean values are plotted below each box.
  }
\label{fig:fig4}
 \vspace{0mm}
\end{figure*}

The boxplots in Fig. \ref{fig:fig4} compare the precision and stability between the ${\bf{x}}_{\sup }$+HT method (setting $H=400$) and GMM+$\Delta {\rm{MFCC}}_{{\rm{sup}}}$ method (setting the number of components as 64).
We can observe that, when $T=50$, the HT model's identification accuracy is a little lower than the traditional GMM model, when the durations of the test utterance data are longer (e.g., $T=100, 150, 200$), the ${\bf{x}}_{\sup }$+HT method can obtain more accurate and stable results.

In order to check the statistical significance of the improvement, we analyzed the statistical independence of these two models by student's $t$-test method. We assumed the identification results from these two models obey independent random normal distributions with equal means and equal but unknown variances. The $p$-value in different $T$ is shown in Table~\ref{table1}, we can observe that when $T=50$, $p$-value is much larger than 0.05 which means statistical independence assumption does not hold. It can be inferred that, when $T=50$, GMM model and HT model have the similar identification effect, although the GMM model achieves a little higher average identification accuracy in $10$ round evaluations. When $T$ is larger than $50$, the $p$-values are much smaller than $0.05$, which indicates the improvement obtained by the HT model over the GMM model is statistically significant.

\begin{table}
\centering
\caption{\small Student's t-test analysis statistical independence between ${\bf{x}}_{\sup }$+HT and $\Delta {\rm{MFCC}}_{{\rm{sup}}}$+GMM method.}
\begin{tabular}{c |c c c c }
\hline
${\bf{\emph{T}}}$ & 50 &100 &150 &200 \\
\hline
${p\text{-value}}$ & 0.1748 & 0.0030 & 0.0158 &0.0193 \\
\hline
\end{tabular}
\ \vspace{0mm}
\label{table1}
\end{table}

Through the above experiments, we can conclude the HT model performs better than the conventionally used GMM model in precision and stability and the HT model can fit the complicated probability distribution better. It encourages us to use the HT model to improve the some other GMM based speech processing system,~\emph{e.g.}, speech recognition system based on the GMM+HMM model.
\section{Conclusions and Further Work}
\label{sec:conclusion}
A speaker identification (SI) method based on histogram transform (HT) model was proposed in this paper. The proposed method used the mel-frequency cepstral coefficients (MFCCs) and the dynamic information among adjacent frames as features. The identification accuracies were improved by using synthesized features generated through the random transform method. By selecting a reasonable number of transforms, more train features were generated to estimate the histogram. The experimental results show that comparing with the traditional GMM model, the HT model make promising improvement for SI tasks.

In the future we can try to use some other features,~\emph{e.g.}, the line spectral frequencies (LSFs) in the HT model. Some other distributions,~\emph{e.g.}, Dirichlet distribution or beta distribution can be used to replace the Gaussian distribution as the prior distribution to estimate the probability of the zero zones of the histogram. Recently, some researches showed that fusion of several different systems effectively improves SI performance\cite{O2013}. Therefore, it is also worthwhile considering fusion of the HT model and the state-of-the-art i-vector based method.


\ifCLASSOPTIONcaptionsoff
  \newpage
\fi

\end{document}